# FAIR: Fuzzy-based Aggregation providing In-network Resilience for real-time Wireless Sensor Networks *

Emiliano De Cristofaro[1], Jens-Matthias Bohli[2], Dirk Westhoff[2]

[1] Information and Computer Science - University of California, Irvine
Irvine, CA, 92617
edecrist@ics.uci.edu
[2] NEC Laboratories Europe
Kurfursten Anlage 36
69115 Heidelberg, Germany
{bohli, dirk.westhoff}@nw.neclab.eu

**Abstract.** This work introduces FAIR, a novel framework for Fuzzy-based Aggregation providing In-network Resilience for Wireless Sensor Networks. FAIR addresses the possibility of malicious aggregator nodes manipulating data. It provides data-integrity based on a trust level of the WSN response and it tolerates link or node failures. Compared to available solutions, it offers a general aggregation model and makes the trust level visible to the querier. We classify the proposed approach as complementary to protocols ensuring resilience against sensor leaf nodes providing faulty data. Thanks to our flexible resilient framework and due to the use of Fuzzy Inference Schemes, we achieve promising results within a short design cycle.

## 1 Introduction

Wireless sensor networks (WSN) are increasingly being used to monitor physical conditions. For an energy-efficient information flow, data aggregation is a core feature of medium and large scale sensor networks. Not only may it be unnecessary to collect the sensed data from each sensor, but it could also result in the nodes' energy rapidly exhausting, due to the high communication load. As a result, most WSNs perform data aggregation – nodes process incoming messages and aggregate the information according to a given aggregation function, e.g. average, sum, minimum/maximum. A subset of nodes, called *aggregator nodes*, build a backbone and are responsible for the data stream. The aggregator nodes may be dynamically selected to balance the load between all nodes.

In this work, we focus on sensor networks that consist of restricted low-cost devices, respond in real-time and possibly without a fixed base station. Nevertheless, the data provided by these wireless sensor networks may have impact on the real world, especially if the data is used to control actuators. For instance, *accident prevention* could be achieved by sensing and providing road conditions: the driver or even the car automatically react on the sensor information at real-time [8]. Moreover, the application of WSNs to SCADA systems could support the protection of critical infrastructures, such as power plants, power distribution networks or oil and gas pipelines [4]. In such settings, we claim the main security goal to be *resilience*, which we characterize by: (i) providing aggregated data *integrity* in the presence of bogus nodes altering the aggregation process, (ii) enforcing *robust* protocols in presence of likely node or message failures, (iii) providing the querier with a *measure on the accuracy* of the aggregated value.

We present FAIR, a novel *resilient* aggregation framework that provides data-integrity, tolerates link or node failures and node misbehavior, and returns a quality measurement of the aggregated value. We argue that the quality measurement is an essential part for robust and secure aggregation. Since failure tolerance can help the adversary to remain undetected with attacks, the integrity protection of the sensor information may be weakened. Therefore, we introduce a quality measure that makes the trust level of the WSN response visible to the querier.

Previous work often concentrates on one protection goal: solutions for (i) are [42,13]. Very recently, protocols that provide integrity and robustness are considered [23,38]. Algorithms that concentrate only on robustness – a solution for (ii) – are [31,29,14,19]. Also recently, robust protocols consider in addition security against corrupted aggregator nodes [20]. Quality of information – as in (iii) – has also been used for data aggregation, e.g. in database systems [32]. WSN solutions providing quality measures are still very rare [9,26]. We are not aware of any protocol that present a solution for (i), (ii) and (iii). More details on related work are provided in Section 6.

---

The FAIR framework uses witness nodes as introduced in [16]. However, FAIR goes one step further and uses multiple witness nodes which not only confirm the result, but aggregate and forward the results themselves. This produces redundancy that can help the protocol to improve data integrity and robustness. Nevertheless, usefully and efficiently utilizing this redundancy is not a trivial task. To this aim, we note that the analysis of all the information available at the aggregator node can be used to infer an estimation on the quality of the aggregated information and choose those witness nodes that provide the highest value. We introduce a value, the *quality of information*, to assess the degree of trust on the aggregated value. More details on our approach is given in Section 3. Our technique is based on the concepts of Fuzzy Inference Schemes (from now on, FIS). Such techniques are known to be effective in making real time decisions using incomplete information [18,21,40]. We claim that FIS are suitable in representing a measure on the quality of information, by means of natural linguistic rules, over input values which cannot be predicted. Moreover, we show that the FIS also suit to handle bogus input values up to a certain degree. We present our Fuzzy Inference Scheme in Section 4 and give an insight on Fuzzy Logic in Appendix A.

We remark that the computation of the quality of information grants a nice plus at the end of the aggregation process, providing the querier with the final measure on the accuracy of the aggregation process. According to the quality of information value the querier decides whether to accept or to decline the result of the query. Note that such a hint on the correctness of the received aggregated value, although still fuzzy, is useful compared to receiving an aggregated value without any quality inspection. Indeed, we show in Section 5 that the proposed framework achieves to improve resilience and provides the querier with a meaningful measure of response accuracy.

## 2   Problem Model

Next, we present an overview about the network assumptions and adversarial models in our work.

**Network Assumptions.** The sensor network consists of $n$ sensor nodes $N_i$, partitioned into clusters. All sensors $N_i$ have a unique identifier $ID_i$. All nodes in a cluster are supposed to be in single-hop communication distance to each other. The clusters do not necessarily have the same shape or size.

The network's functionality is to collect and to provide data. In any epoch, the querier device may query the sensor network for data from this epoch. The owner of the WSN is typically not interested in the entirety of sensor readings, but rather in an application-specific evaluation of the sensor results. We denote the function that derives the result from the sensor readings by $f(v_1, \ldots, v_n)$ where $v_i$ is the sensor reading of sensor $N_i$ in this epoch. To allow data processing within the WSN, it must be possible to decompose the function $f$ into local sub functions, e.g. $f(v_1, \ldots, v_{13}) = f_0(f_1(f_2(v_1, \ldots, v_3), f_2(v_4, v_5), f_2(v_6, v_7)), f_1(f_2(v_8, \ldots, v_{11}), f_2(v_{12}, v_{13})))$. In the rest of this work, we solely assume aggregation functions holding this property. The following protocols are assumed to be in place:

– A routing protocol, e.g. tinyLUNAR [33]. This provides multi-hop communication between sensor nodes. To avoid that forwarding nodes can modify or block messages, a multi-path message propagation is advisable.
– A protocol for authenticated communication between nodes. Nodes have to be able to authenticate messages to other nodes, namely to their aggregator nodes. This protocol could build on the multicast authentication introduced by Canetti et al. [12].
– An aggregator node election protocol. This takes a set of sensor nodes $\{N_1, \ldots, N_n\}$ and the current epoch as input and outputs a random sequence of aggregator nodes $(Agg_1, \ldots, Agg_k) = (N_{i_1}, \ldots, N_{i_k})$, which are then responsible for the aggregation of data of $\{N_1, \ldots, N_n\}$. PANEL [10], LEACH [24], or SANE [37] ensure this. If the node election protocol has to be *non-manipulable* (e.g. fully deterministic), SANE should be used. Furthermore, the protocol needs to be *predictable*, meaning that all inputs are public, such that anyone can compute the aggregator nodes for any epoch and any set of nodes.

**Adversarial Model and Security Requirements.** We assume the adversary is in complete control of the wireless channel. The attacker can eavesdrop data over the wireless broadcast medium or control the communication channel to catch, destroy, modify and send data. As we aim at implementing the protocols on the low-end price class of physically unprotected sensor nodes, the adversary is assumed to be able to take control over several sensor nodes. As a security assumption, a threshold on the number of corrupted nodes will be assumed. If the adversary controls a sensor node, she gains knowledge of all the sensitive information stored at this node. In our security evaluations, we will classify between naïve attacker and smart attackers. The **naïve attacker** operates by means of randomly distributed corrupted nodes and does not perform any kind of collusion or collaboration within the compromised nodes. This adversarial strategy covers a realistic scenario where an attacker manages to cause malfunctions in nodes but does not have a powerful communication equipment in order to let the compromised nodes communicate and collude in an undetectable manner. The attacker's strategy is to perform as much damage as possible during the aggregation process without being discovered by the user. The **smart attacker**'s goal is the same as the naïve – altering the aggregation process without being discovered by the user. However, this attacker is more powerful, in the sense that she cooperatively operates to make the querier accept false data. Compromised witness nodes communicate and collude in order to report the same bogus aggregated value. Moreover, the attacker has some knowledge on the network topology. Thus, she tries to focus the attack on nodes at the same aggregation level, such that witness nodes from the same level can collude. This adversarial strategy covers the worst scenario for our framework.

As formerly discussed, the primary security goals are three. First, we want to protect data integrity and prevent the *stealthy attack* [34] that makes a querier accept a flawed aggregated value. Then, the framework should also be robust, which means the adversary cannot prevent the network from providing a response as long as the number of corrupted nodes and blocked channels is limited. With an increasing number of adversarial interaction, a higher probability of larger deviation of the aggregated value is tolerated and in fact not avoidable. However, the expected deviation should ideally be expressed by a quality of information value (see Section 3.4). Finally, we want to provide the querier with a measure on the accuracy of the entire aggregation process.

## 3 The FAIR Architecture

Witness nodes have been often employed to confirm the result of aggregator nodes, in order to ensure the integrity of data during aggregation. For instance, Du et al. [16] use a witness node that gets the same input as the aggregator node, however, without forwarding the result. Instead, the witness computes a MAC for the result and forwards the MAC to the aggregator node. The aggregator node collects the MACs from the witness nodes and forwards the aggregation result and all MACs to the base station. This approach has been extended to a multi-level aggregator hierarchy in [8]. In a multi-level hierarchy, the confirmation of the witnesses in lower-levels have to be checked within the network. This requires a multicast authentication protocol as outlined in Section 2.

On the contrary, we propose a protocol where several witness nodes not only confirm the aggregator's result, but aggregate and forward the result themselves. Thus, we break down the distinction between the aggregator and witness nodes and will only use the term aggregator nodes (from now on, AN). This results in more robustness and flexibility: there is no dependency on a single AN. Furthermore, the aggregator nodes on a higher level receive the full data and extract information even if the nodes disagree. Figure 1 shows exemplarily an aggregator hierarchy with two aggregating nodes per cluster. All sensor nodes are physically located on the sensor node level, partitioned into clusters. The tree shows a logical aggregator hierarchy. Every node on an in-network aggregation level symbolizes an AN that is elected out of one of its descendants. Thus, it is not a new node, but one of the leaf nodes that fulfills an additional role.

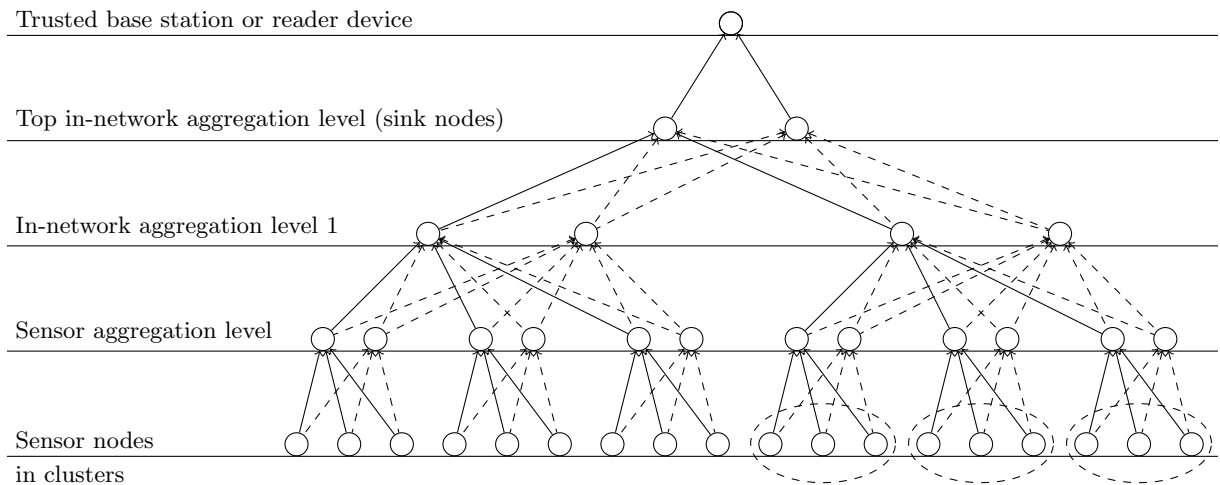

**Fig. 1.** Aggregator hierarchy with two aggregators per cluster. The dashed lines show the communication flow that is caused by the additional aggregator or witness nodes, respectively.

### 3.1 Redundancy

We let the number of aggregators be configurable. This allows the querier to trade resilience for efficiency. As outlined in Section 2, the aggregator node election protocol outputs a sequence of $w$ potential aggregator nodes. Every node in the WSN can potentially act as aggregator node. This comes at the cost of a reduced protection of the data integrity: from the aggregated result, it is not visible if errors or disputes occurred during the aggregation process. However, this gives the flexibility to aggregate via additional nodes in case the assigned aggregator nodes fail or do not reach an agreement. This flexibility helps the protocol robustness, because any aggregator node that fails can be replaced.

### 3.2 Quality of Information

To make anomalies that occurred in the aggregation process visible to the querier, we propose to distributedly compute a value, which we term as Quality of Information (QoI). The QoI conveys the level of trust that can be placed in the aggregated value. The QoI value allows the querier to assess the confidence of the aggregated sensor data. Moreover, the QoI enables an AN to gain the most benefit from the redundancy introduced by the witness nodes. We expect that choosing those witness nodes that provide the highest value will result in a higher accuracy of the aggregation process, despite the presence of adversaries or errors. Certainly, this value is again exposed to attacks and cannot generally be trusted. Here again, the witness architecture is used to validate the QoI value, i.e. a single aggregator that manipulates the QoI gets caught if the responses of further ANs of this subtree are considered.

In principle, it can be distinguished between two origins of quality loss. 1) As discussed above, the resilience introduced to the aggregation process makes a quality processing necessary to account for the increased susceptibility to attacks. 2) Another aspect of QoI arises immediately in the processing of the sensor data on the sensor aggregation level. Here, e.g., the intensity of the event detection or the number of participating sensors can constitute the QoI. The work at hand purely concentrates on the first issue, the processing of the QoI due to anomalies in the aggregation process.

The quality of information will be computed in the data aggregation process. Aggregators on the sensor aggregation level will come up with an initial quality of information value depending on the completeness of sensor data in their cluster. Aggregators in the first in-network aggregation level are the first to start the anomaly based QoI evaluation. Aggregators on higher aggregation levels will continue processing the QoI. The QoI value will be reported to the querier by the sink nodes along with the aggregated WSN information. In Section 4, we will present our implementation computing QoI by means of a Fuzzy Inference Scheme (FIS).

Figure 2 gives a bird's view on the FAIR architecture and highlights its relevant building blocks. An in-network aggregator node, AN is responsible for aggregating messages, e.g., from the subtrees $S_1, S_2, S_3$, where each $S_i$ is the set of sensor nodes at the leafs of the subtree $i$. It receives input data and QoI values from its children and outputs the result of the aggregation and the new QoI value. As presented in Section 3.1, all the nodes in the same subtree should ideally report to AN the same data and QoI values. In reality, however, the presence of attackers or link failures will instead introduce variances. We remark that the use of multiple witness nodes to aggregate the values from the same subset of nodes introduces some redundancy which can now help the protocol resilience. Our idea is to use the QoI value to optimize the selection of nodes within the same subtree from which to get the data. This step is performed by the FAIR inference module, which we introduce in Section 3.3.

We observe that in FAIR, the aggregation algorithm is performed in two steps.

1. The messages belonging to the same subtree are evaluated and filtered by the FAIR inference module. Its output is a value $v_i$ expressing the (aggregated) sensed values of subtree $i$, together with an associated QoI value for the considered subtree in the aggregation hierarchy. The inference module contains the intelligence of the resilient data aggregation protocol, and we will suggest an implementation of this algorithm based on fuzzy logic. We stress that the logic of this module is independent from the concrete aggregation function. More details on this will be shown in Section 5.

2. After the FAIR inference module has been fired, the algorithm *aggregate* performs the actual aggregation function over the results of different subtrees. This algorithm implements a decomposable aggregation function as outlined in Section 2. Moreover, the different QoI values (which have been independently computed in the FAIR inference modules) for each subtree will be averaged to express a unique QoI output as for all the subtrees.

### 3.3 Fuzzy-Inferred QoI

In the following, we look in more detail to the design space of the FAIR inference module. This model harmonizes values received from aggregators within one subtree. The input are data values and QoI estimations of aggregator nodes in a single subtree. As all those aggregator nodes are witnesses to each other, these values are identical in the absence of link failures and attacks.

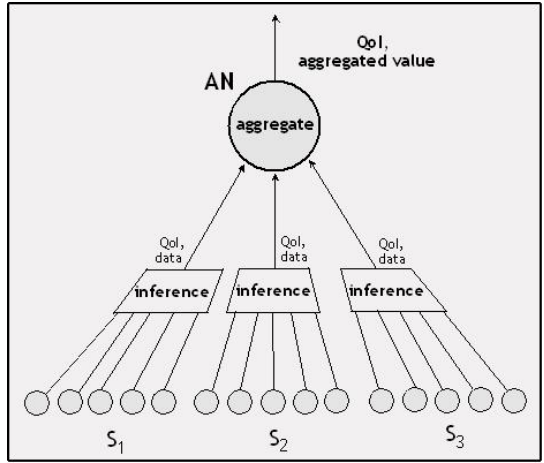

**Fig. 2.** Building blocks of a single aggregator node within the FAIR architecture.

Figure 3 shows the details of the FAIR inference module. In the Figure, we suppose that for each subtree AN is reported with data from 5 witness nodes, but it will consider only 3 of them. However, those parameters are configurable in our implementation. The selection is made by the inference module. Therefore, the inference module has a twofold goal: (i) employing the witness nodes redundancy to use the best selection of nodes from which to get data, and, (ii) employing the QoI to guide the node selection. The inference module consists out of two components: a *fuzzy inference module*, used to compute the QoI and a *filter*, used to select the witness nodes whose data will be further considered in the aggregation process. At the beginning, the *filter* will only select the values received via the radio interface. The *fuzzy inference module* computes the QoI over those inputs and, if this value is above a threshold, then those values will be averaged and the result will be directed to the aggregation function. The computed QoI value serves to estimate the quality of the subtree. However, if the computed QoI value is below the threshold, the node with lowest QoI will be discarded and another node will be considered. This process iterates until the resulting QoI value is above the threshold. If the threshold has not been reached but no further node is available, the process selects the combination of nodes resulting in the highest QoI value.

It is not straightforward to determine on which basis to compute an effective QoI value. In fact, there is no clear indication of ongoing attacks. However, an AN can infer some evaluation on the quality of the aggregated data, by analyzing:

1. *Consistency* - the degree of agreement of lower-level aggregators in the same subtree on the data. A significant deviation indicates the presence of errors or lower-level ANs reporting bogus information.
2. *Completeness* - each AN is aware of the nodes that are expected to send a message (namely, the nodes of the respective subtree $S_i$). If messages of some of the nodes are missing, this could potentially point to link failures or node exhaustion or to an ongoing attack. Also the cases where the message is corrupted or the QoI is so low that the information should actually be ignored can be treated as incomplete information.
3. *QoI from the lower level* - This expresses the QoI evaluation over the data to be aggregated and must be taken into account when computing the current QoI value.

We argue that FISs are beneficial for the FAIR framework. Indeed, FISs are effective in making real time decisions with incomplete information. A potential alternative solution could be to use probabilistic uncertainty models for describing the quality of information. However, those models need significant statistical information to extract probability distributions, but, within this context, there is no such information other than a few vague data. To this aim, we argue that a FIS may provide a possibility to assess QoI by means of natural linguistic terms and rules.

However, we want to point out that the use of a FIS in the FAIR inference module to compute the quality information needs to be defeated against some negative arguments:

- *no setpoint exists to train the module* - this holds except for the ideal case that no attacker is present and all data are transmitted reliably, for which we should have perfect QoI;

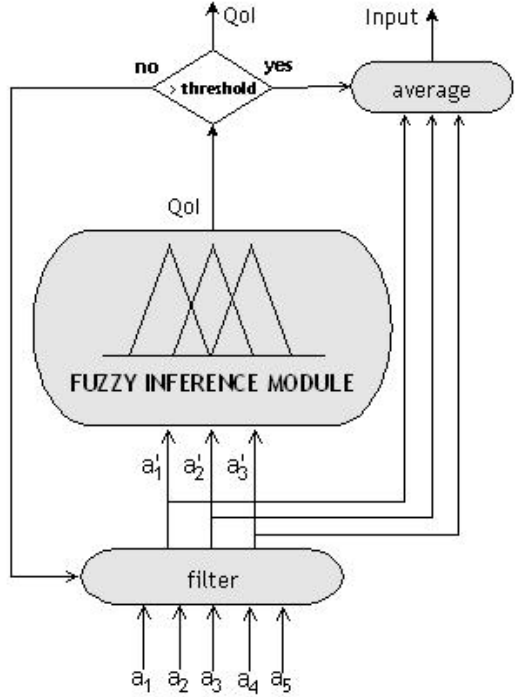

**Fig. 3.** The FAIR inference module.

— *no feedback loop exists* - generally fuzzy controllers continuously fire rules with the aim to stabilize a process by comparing its actual output value with the setpoint; in our setting for each aggregation process several fuzzy inference module fire once without any feedback loop for a given setpoint;
— *distributed* - typical fuzzy controllers are centralized whereas our scheme is highly decentralized, i.e. the same fuzzy inference module instantiation is located on each aggregator nodes firing probably different rules with different actual input values for QoI, Completeness and Consistency;
— *bogus data* - Fuzzy controllers are usually applied to control loops where the actual status of the process differs from the setpoint due to environmental influences and/or overmodulation of the actual output value of the controller, and to due to adversaries;

### 3.4 Remarks on the interpretation of QoI

At the end of the aggregation process, the sink node will report the querier both the final aggregated value and the final QoI value, which would ideally express the degree of trust which can be put on the response accuracy. For this value, we propose the following semantic:

$$\text{QoI}^{out}\begin{cases} \text{do not use if } \text{QoI}^{out} \leq 0.5 \\ \text{only in non-sensitive settings if } 0.5 < \text{QoI}^{out} \leq 0.8 \\ \text{use with high confidence otherwise} \end{cases}$$

Such a coarse granular classification is easily understandable and therefore applicable by the querier. However, we want to point out some observations with respect to this classification:

For a $\text{QoI}^{out}$ belonging to the category 'use with high confidence' the results may have different causes:

1. The system is running in the absence of any bogus modifications without considerable link failures.
2. Some minor malicious nodes may have incorporated bogus data. However, these data with a high probability have been overruled by the FAIR distributed fuzzy inference module; this is true for the naive attacker as long as the corrupted nodes are relatively remote from the sink node.

3. A majority of bogus nodes from the same subtree agrees on the same wrong value. According to the concrete level in the aggregation hierarchy, the impact is different.

As a result, in case the QoI$^{out}$ is ranked as 'use with high confidence' although there is still some uncertainty that bogus data have effected the output, we will see that to a high degree one can trust the aggregated value.

Vice versa, for a QoI$^{out}$ belonging to the category 'do not use' the results are out of the following reasons:

1. There was no attack; however, for one or the other reason not enough nodes contributed.
2. There was an attack and/or not enough nodes contributed such that the aggregation process is not presentable.

To conclude, in the case QoI$^{out}$ is ranked as 'do not use' both subcases clearly indicate that one should not use the aggregated node. Vice versa, 'use with high confidence' means that data can be trusted to a high degree. Here only case 3 could cause the querier to accept a bogus aggregated value despite a high QoI, i.e. there is a colluding attacker.

## 4 Fuzzy Inference Module

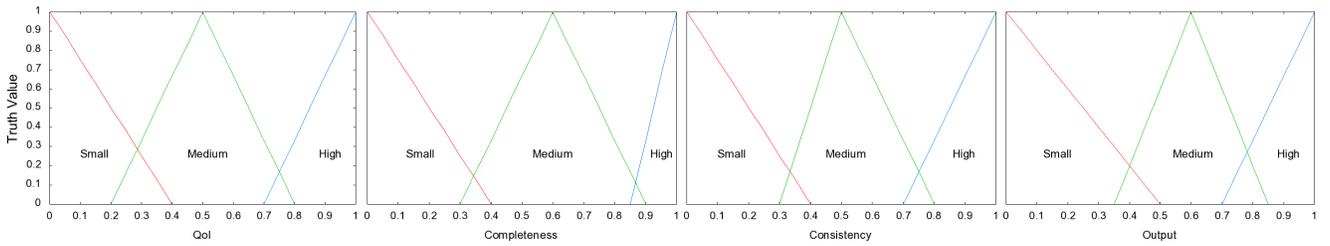

**Fig. 4.** Fuzzy Sets describing the Fuzzy Input and Output Variables.

### 4.1 Design

At each aggregation step, a measure on the quality of information is computed by the aggregator nodes to assess the level of trust that can be placed on the aggregated value. Our approach is based on Fuzzy Inference Schemes (FIS). Over the last years, the research community has proved that such schemes are very handy for making real time decisions with incomplete information. Standard control systems in fact need to rely on an accurate representation of the environment. By contrast, FISs are suitable in representing a measure on the quality of information, by means of natural linguistic rules, over input values which are imprecise and potentially bogus.

Recall that our lightweight implementation neither requires any change to the network topology nor does it require any central authority to execute the fuzzy inference module. In fact, each aggregator node runs the same instance of the controller with its local inputs.

**Fuzzy Variables.** At each step, an AN computes the QoI using these three fuzzy variables:

1. *Completeness* - the fraction of children nodes responding to the query.
2. *Consistency* - normalized standard deviation on data received from children nodes.
3. QoI- the QoI values received from the children nodes.

**Fuzzy Sets.** The linguistic variables used to represent the fuzzy variables are intuitively and simple. We distinguish among three attributes: small, medium, and high. However, for each fuzzy variable those attribute have a different meaning and may have a slightly different shape according to our empirical experience. Figure 4 shows the different fuzzy sets for the three input fuzzy variables and for the fuzzy output.

**Fuzzy Rules.** The core part of every fuzzy controller is the choice of the set of rules. In a first stage, we define a very limited set of disjoint rules to mirror the straightforward expectation on the evolution of quality of information values over the aggregation process. We define the following set of rules:

(a) IF (QoI is small AND Consistency is small AND Completeness is small) THEN QoI$^{out}$ is small

(b) IF (QoI is medium AND Consistency is medium AND Completeness is medium) THEN QoI$^{out}$ is medium

(c) IF (QoI is high AND Consistency is high AND Completeness is high) THEN QoI$^{out}$ is high

Subsequently, we iteratively refine our set of rules in order to capture a greater number of possible scenarios. In fact, the above rules do not catch the conditions where the inputs have different fuzzy attributes. First, we remark that the quality of information value should be computed as small not only when all the three inputs are small. In the spirit of a conservative risk assessment, one small input value represents with high probability an ongoing attack. Therefore, we skip rule (a) and add the following rules:

(1) IF QoI is small THEN QoI$^{out}$ is small

(2) IF Completeness is small THEN QoI$^{out}$ is small

(3) IF Consistency is small THEN QoI$^{out}$ is small

Furthermore, we note that ideally the quality of information should be computed as high only when all the three input values are high. As a result, the following rules are added to the rule set to have the output to medium when at least one input value is medium:

(4) IF (QoI is medium AND Consistency is medium AND Completeness is high) THEN QoI$^{out}$ is medium

(5) IF (QoI is medium AND Consistency is high AND Completeness is medium) THEN QoI$^{out}$ is medium

(6) IF (QoI is medium AND Consistency is high AND Completeness is high) THEN QoI$^{out}$ is medium

(7) IF (QoI is high AND Consistency is medium AND Completeness is medium) THEN QoI$^{out}$ is medium

(8) IF (QoI is high AND Consistency is medium AND Completeness is high) THEN QoI$^{out}$ is medium

(9) IF (QoI is high AND Consistency is high AND Completeness is medium) THEN QoI$^{out}$ is medium

Moreover, we remark that having all three inputs as medium mirrors w.h.p. an ongoing attack as well as the impossibility for the fuzzy inference module to recover the aggregation process. Hence, we substitute rule (b) with:

(10) IF (QoI is medium AND Consistency is medium AND Completeness is medium) THEN QoI$^{out}$ is small

Finally, we save rules (c):

(11) IF (QoI is high AND Consistency is high AND Completeness is high) THEN QoI$^{out}$ is high

Finally, we choose the *min-max* inference scheme and the weighted average method for the defuzzification. For more details, we refer to Appendix A.

We remark that the FAIR inference module is deployed into each AN performing in-network aggregation (see Figure 1). In fact, at the sensor aggregation level, the only information available to the aggregator nodes is the Completeness. These nodes do not receive any quality of information from the child nodes and they cannot compute any Consistency measure from the sensing nodes.

## 4.2 Implementation

We distinguish two different stages of the implementation of the fuzzy inference module: (i) **testing** and (ii) **deployment**.

The testing phase is essentially needed to test the behavior of the module in different network's and adversarial settings. During this phase, we work in a Java environment, enhanced by the JGraphT library [1] and the FuzzyJ toolkit [6]. The JGraphT package is a free Java graph library that provides mathematical graph-theory objects and algorithms. We use this package to simulate a sensor network through an undirected graph. Sensor nodes are simulated by nodes in the graph. To represent two nodes within their transmission range, an edge is inserted in the graph. The FuzzyJ toolkit instead is a set of Java classes that provide the capability for handling fuzzy concepts and fuzzy inference based on fuzzy logic. We use this package to implement the fuzzy inference module. At this stage, we design a simple and flexible network model and run simulations in order to tune the fuzzy sets and rules, as for the rules and the fuzzy sets.

In a second stage, we focus on how to deploy our framework on real sensor nodes, e.g. Mica2/MicaZ. We are aware that sensor nodes are tiny devices with limited memory, computational power, and battery life. As a result, the introduction of the Fuzzy Controller has to be carefully designed. It is in fact unfeasible to embed a complex library in a sensor node to implement the Controller. Therefore, we make use of the JFS toolkit [30]. JFS is a development environment for the programming language JFL. The environment includes tools to compile, run, improve and convert JFL programs. JFL is a special-purpose language used to write functions. It combines features from traditional programming languages like (C, Pascal, Basic, etc.) with fuzzy logic and machine learning. The most important feature of this toolkit for our goal is the ability to convert the JFL code to a source code file in standard ANSI C. The resulting source code does not need any additional library or package, but it is ready to run. Furthermore, the resulting algorithm does not require any relevant computational power, since in practice the fuzzy inference module will only make a set of function evaluations and cascaded if-then-else statements. Having standard and simple C code allows the deployment on a large set of sensor network platforms. Indeed, the most diffused types of sensor nodes, i.e. Mica motes run the TinyOS operating system [7]. This allows to easily deploy algorithms written in the nesC programming language [3], an extension to the C programming language designed to embody the structuring concepts and execution model of TinyOS. Hence, our final step is to use the JFS toolkit to generate a standard C source code for our fuzzy inference module. Then, porting the C code to nesC is straightforward and we can finally deploy our fuzzy-driven aggregation mechanism on Mica motes.

## 5   Security and Overhead Analysis

In this section, we evaluate our framework. We want to show that the use of Fuzzy-inferred Quality of Information improves the accuracy of the final aggregated value, (i) in presence of bogus nodes altering the aggregation process, (ii) in presence of node or link failures. Moreover, (iii) we show that our framework provides the querier with a reasonably accurate estimation of the quality of the query process, so that the querier has a rationale to decide whether to accept the result or not. In this section, we show that the use of the fuzzy inference module remarkably increases the accuracy of the aggregation process, without performing any heavy operation or increasing the communication load. Indeed, the use of the fuzzy inference module embedded into the FAIR architecture helps the framework improving the choice of the witness nodes used to compute the aggregation function. The presence of bogus ANs results in fact in lower Consistency values, as well as the failure of some links results in lower Completeness values, and thus lower QoI for the subtree. This would lead upper level ANs to try to discard the aggregated value coming from that subtree in favor of subtrees with higher QoI. On the contrary, without using the fuzzy inference module, the AN would have no way to perform this choice.

To confirm our analysis, we perform an simulation analysis using the Java testing environment presented in Section 4.2. As presented in Section 2, we distinguish between two types of attackers: (i) the *naïve attacker* compromises bogus nodes randomly distributed within the network, and (ii) the *smart attacker* coordinates colluding bogus nodes.

**Network topology.** We used a reasonable average sized sensor network, composed by 100 sensing nodes. We choose an average cluster size of 12 nodes and we set ANs at the in-network aggregation levels to aggregate from an average of 3 lower level ANs. Finally, we choose an average number of 4 witness nodes for each AN. As a result, we have 3 aggregation levels, plus the final one at the sink.

Finally, we present an overview on the energy and computation overhead needed by our framework.

### 5.1   Naïve Attacker

In a first stage, we evaluate the behavior of our framework in presence of bogus ANs, assuming neither node nor link failures. We let a bogus node report a value randomly ranging from the 10% to the 90% of the correct value. The chosen aggregation function is the *average*. Figure 5(a) shows a remarkable increase in the accuracy thanks to the use of the fuzzy inference module, enabling every AN to improve the choice of witness nodes.

Subsequently, we consider the case where, besides bogus ANs, a percentage of links fail. This could happen for two possible reasons: (i) the ANs has prematurely run out of battery and is unable to receive/transmit messages, (ii) an adversary is occupying the node and preventing some or all message transmissions. For this experiment, we use the same network configuration. However, in this case instead of bogus ANs, we simulate a ranging number of link failures. Link failures themselves do not falsify the result. The lower quality is justified by the reduced redundancy that facilitates attacks. Therefore, we simulate the settings where both link failures and a fixed 10% of bogus nodes occur. Figure 5(b) confirms a remarkable gain in the accuracy thanks to the use of the the fuzzy inference module also with this setting. Next, we show that in case of a naïve attack the final QoI value received by the querier is effectively providing a reasonable measurement on the degree of trust to give the aggregation process, according to the semantic introduced in Section 4. For this reason, we run some simulations showing that w.h.p. the QoI$^{out}$ value corresponds to the degree of accuracy of the final aggregated value. Once again, the average aggregation function has been chosen. Figure 5(c) shows that the accuracy of the aggregation process grows linearly with the QoI value. Furthermore, it shows that to each class of QoI values (do not use, use only in non-sensitive settings, use with high confidence) corresponds the right degree of query accuracy. We conclude that in the presence of purely naïve attackers, our QoI-based metric can be effectively used to evaluate the entire aggregation process at real time, without the need of any post-aggregation communication, such as commitment disclosure (as in [42,13]).

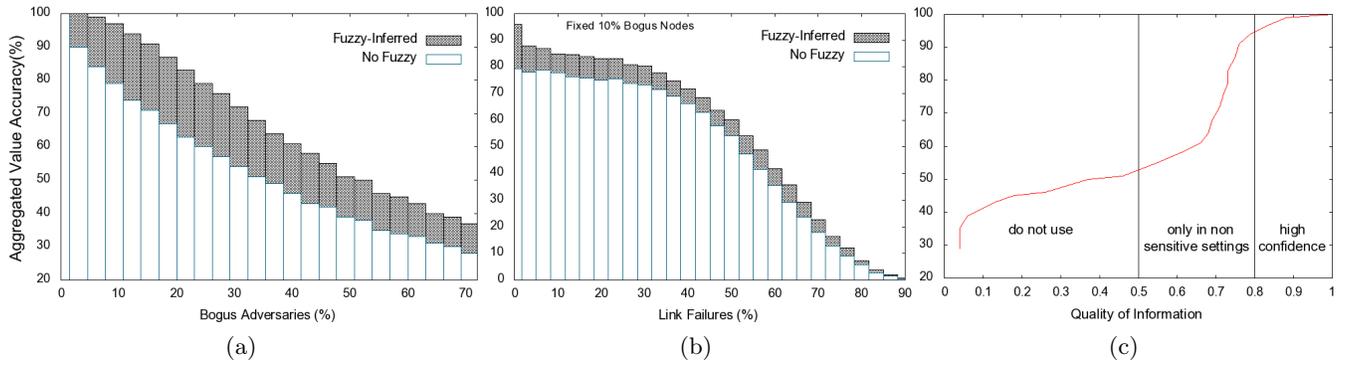

**Fig. 5.** Naïve adversary

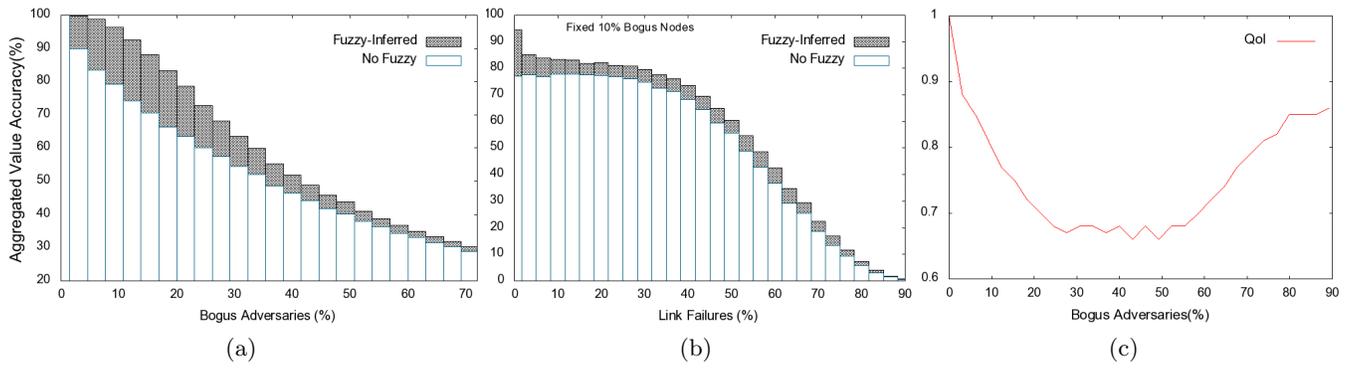

**Fig. 6.** Smart adversary without knowledge of the topology

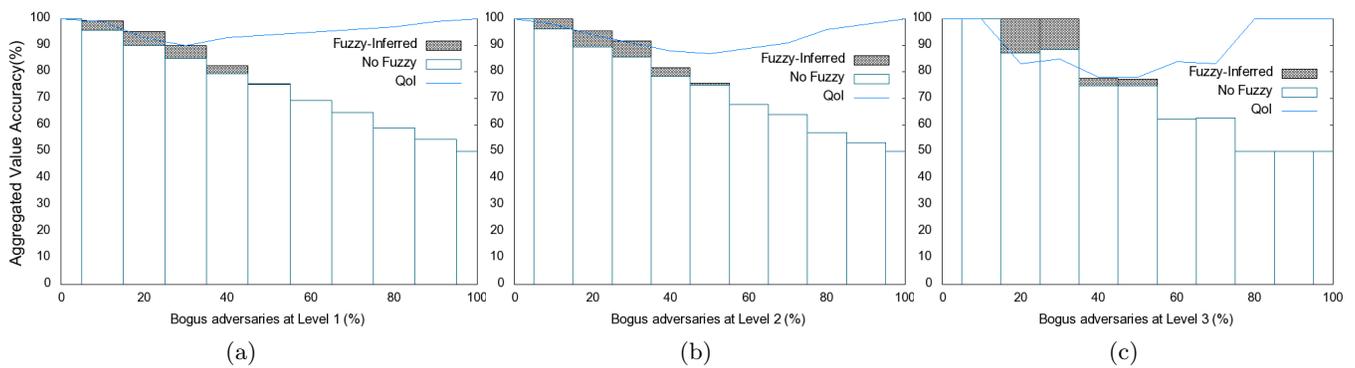

**Fig. 7.** Smart adversary with topology knowledge

## 5.2 Smart Attacker

We perform the same set of experiments as for the naïve attacker with the same topology, and with the same aggregation function, i.e. the average. In a first stage, we let the adversary collude, i.e. report the same bogus value when acting as a witness for the same subtree. All the adversaries report the 50% of the correct aggregate value.

Figure 6(a) focuses on the presence of bogus ANs, while Figure 6(b) on the presence of 10% of bogus ANs and an increasing number of link failures. The gain obtained by the use of the FAIR inference module is less in comparison with the naïve adversary, but still significant. Finally, Figure 6(c) gives the idea to how degree the QoI reflects reality in presence of smart adversary. As we can see, after about 30% of node compromission, the inference module is fooled by the collusion of the adversaries. In fact, the QoI value begins to increase instead of decreasing w.r.t. the presence of bogus adversaries. As a result, we remark that the FAIR framework can be considered resilient in presence of no more than approximately 30% of compromised nodes.

In a second step, we test the situation where the adversary is also given information on the topology, such that she can focus on attacking nodes at the same level, besides colluding to report the same bogus information. The network used for our simulations is still the same as presented above. Therefore, under this setting three different aggregation levels are presented and attackable.

Figure 7 shows only a slight accuracy gain through the use of the FAIR inference module on the three different aggregation levels. Furthermore, because of the collusion, an adversary compromising ANs at the same level can fool the inference module which will always obtain a high level of consistency and completeness. As a result, in this extreme setting, the QoI provides no remarkable information on the accuracy of the aggregation process. However, we argue that cryptographical protocols addressing this setting will require a high number of messages exchange (and thus they are not practicable for real-time traffic).

## 5.3 Independence from aggregation function

In the previous sections, we evaluated the behavior of the framework when the *average* is chosen as aggregation function. This function has the nice collateral feature of mitigating errors by averaging multiple values. In the following, we want to highlight that the resilience of our framework does not depend on the particular used aggregation function. Therefore, we present some experimental evaluation on the framework with not-mitigating functions such as the maximum computation (from now on *max*).

The results have the same degree of accuracy and significance of the ones showed for the *average* aggregation. Figure 8 shows the results for a naïve attacker for *max* aggregation, while Figure 9 focuses on a smart attacker without topology knowledge for *max* aggregation. Recall that all the adversaries report the 50% of the correct aggregate value. knowledge. Finally, Figure 10 tests the situation where the adversary is also given information on the topology, such that she can focus on attacking nodes at the same level, besides colluding to report the same bogus information. The network used for our simulations is still the same as introduced. However, the aggregation function is the *max*. These results impressively indicate that the FAIR framework provides in-network resilience independently of the actual aggregation function.

## 5.4 Overhead

Finally, we want to show that the use of the fuzzy inference module (as the core component of our FAIR architecture) to drive the aggregation process results in no relevant time or energy overhead. After deploying our code in the TinyOS environment, we evaluate the energy and computation overhead by means of the Avrora Simulation Framework [5]. This tool is a set of simulation and analysis tools for programs written for the AVR microcontroller produced by Atmel and the Mica2 sensor nodes. In particular, we test our code on the Mica2dot motes, which are smaller than Mica2 ones and better suited for commercial deployment [2]. We remark that we evaluate only the FAIR inference module, not the entire framework aggregation architecture.

The data transmission costs for the QoI (4 or 8 bits) do not cause any additional energy consumption for transmission. The footprint, i.e. the occupied memory, is as low as 36KB. The simulations show that our FAIR inference module employs an average amount of $4 \cdot 10^{-4}$ Joule and about 13000 CPU cycles, which took an average time of about 1ms for firing the rules. Note that this is even more efficient than a hash function computation. As a result, we claim that the use of the fuzzy inference module to drive the aggregation process results in no remarkable overhead, whereas it improves the resilience of the aggregation process without the use of any heavy cryptographical operation.

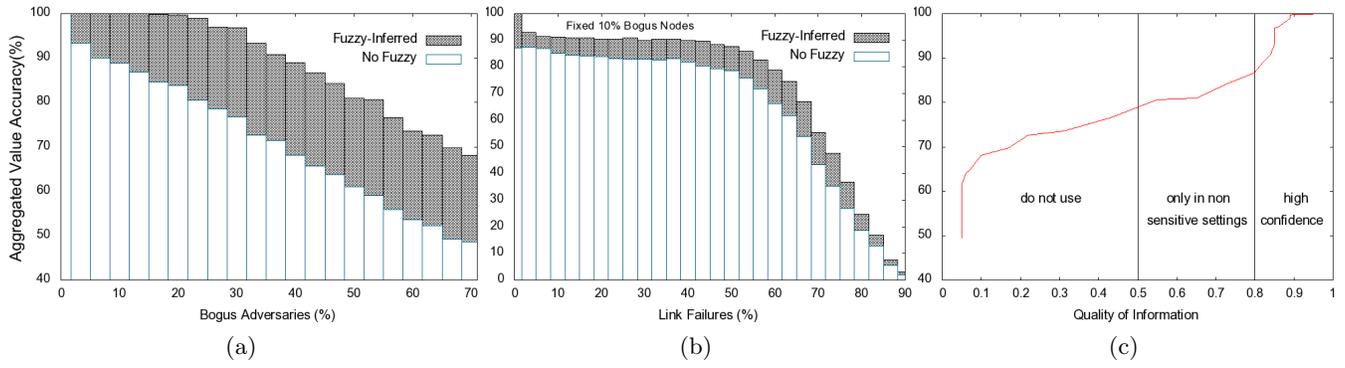

**Fig. 8.** Naïve adversary for *max* aggregation

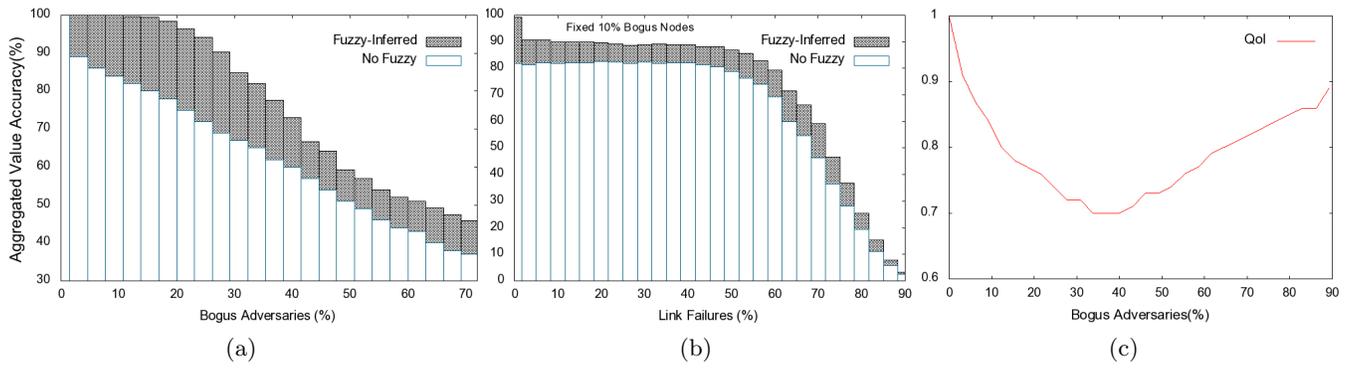

**Fig. 9.** Smart adversary for *max* aggregation without topology knowledge

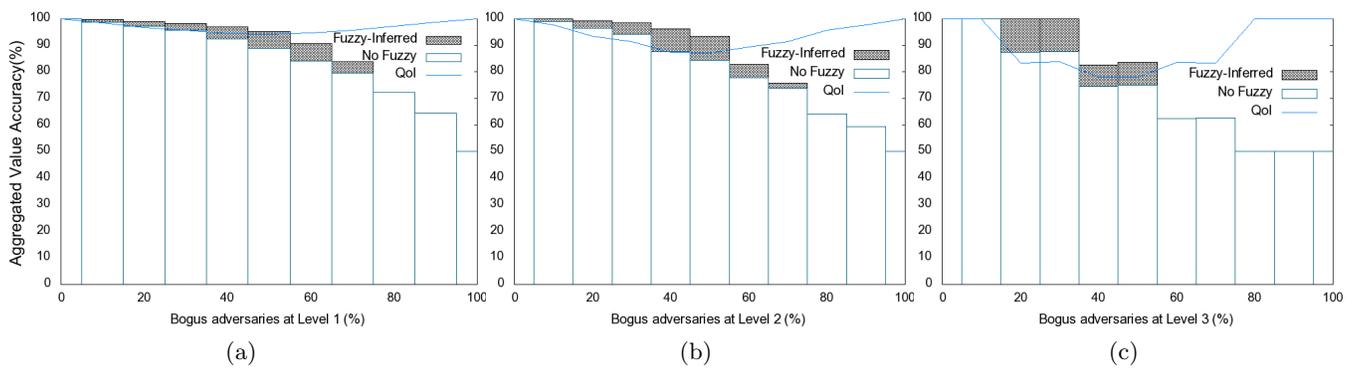

**Fig. 10.** Smart adversary for *max* aggregation with topology knowledge

## 6 Related Work

Faulty data can originate from a sensor node or an malicious aggregator. As the adversary is always able to manipulate the sensed environment such that a sensor will report faulty data, the aggregation function should limit the influence of a single sensor. Such proposals can be found in e.g. [39,11]. Protection against malicious aggregation is guaranteed by protocols such as [42,13]. The protocols first aggregates the information without ensuring integrity. In a second step, the reader initiates the validation of the information. Thus, we argue that the protocols are not well suited for unattended sensor networks where a reader device or base station is only temporarily present [15]. The protocols do also not tolerate any failure, only recent proposals of Haghani et al. [23] and Taban et al. [38] bring in capabilities to detect malicious nodes as a way to provide robustness. SecureDAV [28] uses threshold public key cryptography, which we argue to be not suitable in a sensor network. Another way, that is followed by FAIR, is to base the integrity on witness nodes [16]. Robustness can be provided by requiring only a threshold witnesses to answer.

Protocols that aim at robust aggregation do frequently not use any aggregation hierarchy but use directed flooding to spread the messages [31,29,14]. This creates multiple aggregation paths and needs the aggregation function to be invariant against double counting. The latter can be achieved by approximate algorithms using counting sketches [19]. Recent proposals also consider approximate aggregation that is secure against corrupted aggregator nodes [20].

The idea of embedding into data additional details, such as *quality of information*, is a common solution in many problems, such as knowledge discovery or data mining. The same concept has been used in the context of data aggregation. For instance, the well-known solution proposed in [32] aimed to "drive" the integration of heterogenous databases by exploiting quality-information embedded in the data. Furthermore, in the context of data aggregation in WSN, additional quality information has been used to enhance resilience of computation. In [9], nodes estimate the quality they can deliver. In [26], authors proposed an aggregation framework that can detect and isolate spurious measurements from computed aggregate values, by adding feedback information to the data.

The applicability of Fuzzy Control Systems to security problems has been first proposed in [25], where the notion of failure was represented in terms of fuzzy sets and widely interpreted. Several application aspects of fuzzy methodology in system failure engineering were discussed. Instead, authors in [35] introduced a mechanism to control QoS in IP Networks using Fuzzy Logic. A Fuzzy Controller is implemented in network routers in order to evaluate policy conditions. For instance, the Fuzzy Controller takes as input information such as link status, packet loss, delay, jitter, and so on. It then generates the policy actions as Fuzzy outputs, i.e. values needed for traffic classifications, traffic shaping, dropping, scheduling, and so on.

Fuzzy Controllers have also been used in the context of WSN. The work in [22] introduced a Fuzzy-based approach for the problem of cluster-head election. The Fuzzy Controller is performed by a central control algorithm run at the base station, which is assumed to have global knowledge over the network). As input, the controller takes three Fuzzy values: energy, concentration, and centrality. The output is basically a probability for any single node to become cluster-head. Finally, in [27], a Fuzzy-based mechanism to face *selective forwarding* attacks – malicious nodes dropping sensitive packets – in WSN. In particular, it addresses this problem by letting nodes select multiple paths in order to deliver data. However, the number of different paths must be optimized in order to achieve the best compromise between security and efficiency. For this reasons, authors make use of a Fuzzy Controller running at the base station, which takes as input the energy information and an estimation of the number of malicious nodes, and gives as a fuzzy output the suggested number of paths.

## 7 Conclusions

We proposed the Fuzzy-based FAIR framework for resilient data aggregation in real-time responsive wireless sensor networks supporting in-network processing. Our simulation results validate that a distributively computed QoI value provides a remarkably good measure according to the accuracy of an aggregated value. Moreover, we show that the use of the inference scheme improves the accuracy of the aggregation process. This is in particular true for naïve attackers and to some degree also holds for smart attackers. We highlight that the FAIR framework suits for different classes of aggregation functions like average and min/max functions. Finally, being stimulated by our promising results based on Fuzzy Inference Schemes considering different classes of attackers, we advocate the security research community to make broader use of such concepts.

# A   Appendix: Fuzzy Control Systems

Fuzzy Logic concerns a fundamental methodology to represent and process linguistic information, with mechanisms to deal with uncertainty and imprecision [43]. It extends boolean logic to assign to each proposition a *degree of truth* value in between 0 and 1. As a result, the so-called *fuzzy sets* are sets whose elements have *degrees of membership*. In classical set theory, the membership of elements in a set is expressed by a binary value according to a bivalent condition – an element either belongs or does not belong to the set. By contrast, fuzzy set theory allows the gradual assessment of the membership of elements in a set; this is described with the aid of a membership function (denoted by $\mu$), valued in the real unit interval $[0, 1]$. Fuzzy sets generalize classical sets, since the indicator functions of classical sets are special cases of the membership functions of fuzzy sets, if the latter only take values 0 or 1[17]. The most common shape of membership functions is triangular, although trapezoidal and bell curves are also used. The process of converting a set of real (*crisp*) numbers onto fuzzy membership values is called *fuzzification*. An example of such process is given in Figure 11.

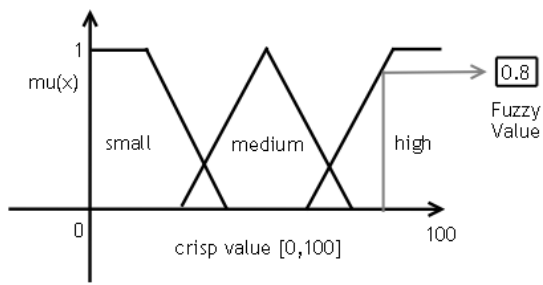

**Fig. 11.** Fuzzification

Fuzzy Control Systems (FCSs) are based on fuzzy logic. Input variables in a FCS define the language used to discuss a *fuzzy* concept. The *fuzzy variables* are characterized by a name (e.g., temperature), a universe of discourse (e.g., $[0-100]$), the units of the variable (e.g., centigrade), and a set of primary *fuzzy terms* (e.g., `hot`, `cold`, `warm`). Terms can also have modifiers (e.g., `very`) and operators (e.g., `and`). As for standard control systems, we distinguish between *input* and *output* variables. In FCSs, each fuzzy variable will have an associated fuzzy membership value.

The key element of a FCS is the set of *fuzzy rules*. These govern and assess the behavior of the system, generating a fuzzy output starting from the fuzzy input values. Fuzzy rules are sets of statements in the form of `if antecedent then conclusion`. Every rule is evaluated in parallel using fuzzy reasoning. All the rules that apply are invoked, using the membership functions and truth values obtained from the inputs, to determine the result of the rule. There are several methods to determine the result. However, the most common one is the *max-min inference scheme* [41], in which the output membership function is given the truth value generated by the premise (min). Then, the results of all the rules that have fired are combined, by performing a *fuzzy union* of the membership values (max). This result in turn needs to be converted into a concrete (crisp) value, through a procedure known as *defuzzification*. There are several different methods to defuzzify a value, each with various advantages and drawbacks. Among the others, the *center of gravity* method is very popular: the center of mass of the result provides the crisp value, and it is computed by the following formula:

$$\texttt{COG} \ = \frac{\displaystyle\int \mu_i(x) \cdot x \ dx}{\displaystyle\int \mu_i(x) \ dx} \tag{1}$$

where $\mu_i(x)$ is the aggregated membership function and $x$ the output fuzzy variable. Another approach is the *weighted average* method, which finds the weighted average of the x values of the points that define a fuzzy set using the membership values of the points as the weights, according to the following formula:

$$\texttt{W\_AVG} \ = \sum_{i=1}^{n} m^i w_i \Big/ \sum_{i=1}^{n} m^i \tag{2}$$

where $m^i$ is the membership function of each rule, and $w_i$ is the weight associated with each rule. This method is computationally faster and easier, and gives fairly accurate result [36]. An example of a FCS using the max-min inference scheme is given in Figure 12.

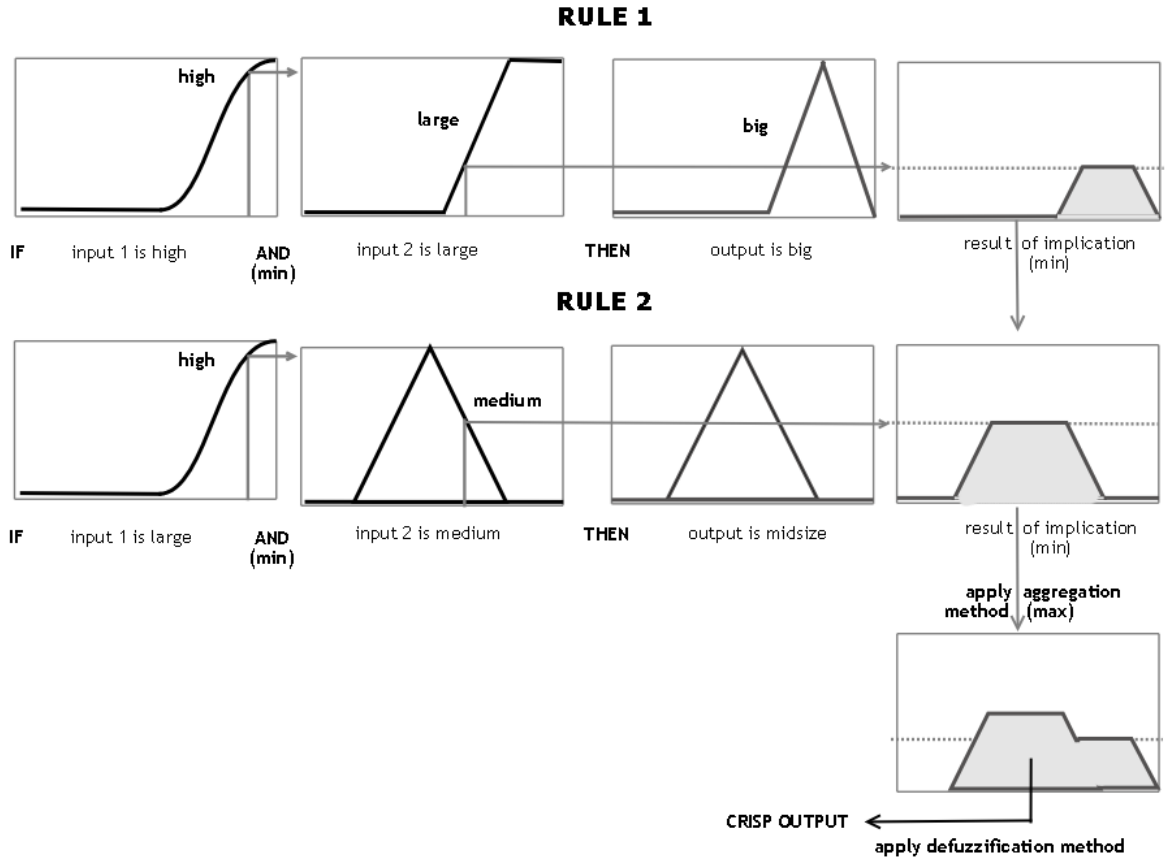

**Fig. 12.** Max-min Inference Scheme